\pdfoutput=1

\documentclass[11pt,twoside,a4paper,cmspaper,final,collab]{cms-tdr}
\def\svnVersion{471759P}\def\cmsCernNoTag{CERN-EP-2018-061}\def\cmsCernDate{\today}\def\cmsMessage{Published in the Journal of High Energy Physics as \href{http://dx.doi.org/10.1007/JHEP08(2018)016}{\doi{10.1007/JHEP08(2018)016.}}}

\begin{document}\cmsNoteHeader{EXO-16-044}

\hyphenation{had-ron-i-za-tion}
\hyphenation{cal-or-i-me-ter}
\hyphenation{de-vices}
\RCS$HeadURL: svn+ssh://svn.cern.ch/reps/tdr2/papers/EXO-16-044/trunk/EXO-16-044.tex $
\RCS$Id: EXO-16-044.tex 471755 2018-08-13 20:39:08Z ahart $

\newlength\cmsTabSkip\setlength{\cmsTabSkip}{1ex}
\providecommand{\cmsLeft}{left\xspace}
\providecommand{\cmsRight}{right\xspace}

\newcommand{\dz}{\ensuremath{d_0}\xspace}
\newcommand{\MT}{\ensuremath{m_{\mathrm{T}}}\xspace}
\newcommand{\MZ}{\ensuremath{m_{\mathrm{Z}}}\xspace}
\newcommand{\PSGcmpDo}{\ensuremath{\widetilde{\chi}^\mp_{1}}\xspace}
\newcommand{\Mll}{\ensuremath{m_{\ell\ell}}\xspace}
\newcommand{\Ztoll}{\ensuremath{\cPZ\to\ell\ell}\xspace}
\newcommand{\Ztoee}{\ensuremath{\cPZ\to\Pe\Pe}\xspace}
\newcommand{\Ztomumu}{\ensuremath{\cPZ\to\Pgm\Pgm}\xspace}
\newcommand{\Ztotautau}{\ensuremath{\cPZ\to\Pgt\Pgt}\xspace}
\newcommand{\Wtolnu}{\ensuremath{\PW\to\ell\nu}\xspace}
\newcommand{\Pveto}{\ensuremath{P_{\text{veto}}}\xspace}
\newcommand{\Poffline}{\ensuremath{P_{\text{offline}}}\xspace}
\newcommand{\Ptrigger}{\ensuremath{P_{\text{trigger}}}\xspace}
\newcommand{\Pfake}{\ensuremath{P_{\text{spurious}}}\xspace}
\newcommand{\Nctrll}{\ensuremath{N_{\text{ctrl}}^{\ell}}\xspace}
\newcommand{\NctrlBasic}{\ensuremath{N_{\text{ctrl}}^{\text{basic}}}\xspace}
\newcommand{\Nestl}{\ensuremath{N_{\text{est}}^{\ell}}\xspace}
\newcommand{\NestFake}{\ensuremath{N_{\text{est}}^{\text{spurious}}}\xspace}
\newcommand{\Ntp}{\ensuremath{N_{\text{T\&P}}}\xspace}
\newcommand{\NtpVeto}{\ensuremath{N_{\text{T\&P}}^{\text{veto}}}\xspace}
\newcommand{\NSStp}{\ensuremath{N_{\text{SS\,T\&P}}}\xspace}
\newcommand{\NSStpVeto}{\ensuremath{N_{\text{SS\,T\&P}}^{\text{veto}}}\xspace}
\newcommand{\metWithMu}{\ensuremath{p_{\mathrm{T}}^{\text{miss},\,\Pgm}}\xspace}
\newcommand{\metWithMuVec}{\ensuremath{\vec{p}_{\mathrm{T}}^{\kern1pt\text{miss},\,\Pgm}}\xspace}
\newcommand{\metMinusOneVec}{\ensuremath{\vec{p}_{\mathrm{T}}^{\kern1pt\text{miss}} + \vec{p}_{\mathrm{T}}^{\kern1pt\text{lepton}}}\xspace}
\newcommand{\metMinusOneUpVec}{\ensuremath{\vec{p}_{\mathrm{T}}^{\kern1pt\text{miss}} + \frac{p_{\mathrm{T}}^{\text{lepton}} - 10\GeV}{p_{\mathrm{T}}^{\text{lepton}}} \vec{p}_{\mathrm{T}}^{\kern1pt\text{lepton}}}\xspace}
\newcommand{\Ecalo}{\ensuremath{E_{\text{calo}}}\xspace}
\newcommand{\dPhiJetMET}{\ensuremath{\abs{\Delta\phi (\text{leading\;jet},\,\ptvecmiss)}}\xspace}

\cmsNoteHeader{EXO-16-044}
\title{Search for disappearing tracks as a signature of new long-lived particles in proton-proton collisions at $\sqrt{s} = 13\TeV$}

\date{\today}

\abstract{A search is presented for long-lived charged particles that decay within the CMS detector and produce the signature of a disappearing track. A disappearing track is an isolated track with missing hits in the outer layers of the silicon tracker, little or no energy in associated calorimeter deposits, and no associated hits in the muon detectors. This search uses data collected with the CMS detector in 2015 and 2016 from proton-proton collisions at a center-of-mass energy of 13\TeV at the LHC, corresponding to an integrated luminosity of 38.4\fbinv. The results of the search are interpreted in the context of the anomaly-mediated supersymmetry breaking model. The data are consistent with the background-only hypothesis. Limits are set on the product of the cross section for direct production of charginos and their branching fraction to a neutralino and a pion, as a function of the chargino mass and lifetime. At 95\% confidence level, charginos with masses below 715 (695)\GeV are excluded for a lifetime of 3 (7)\unit{ns}, as are charginos with lifetimes from 0.5 to 60\unit{ns} for a mass of 505\GeV. These are the most stringent limits using a disappearing track signature on this signal model for chargino lifetimes above $\approx$0.7\unit{ns}.}

\hypersetup{
pdfauthor={CMS Collaboration},
pdftitle={Search for disappearing tracks as a signature of new long-lived particles in proton-proton collisions at sqrt(s) = 13 TeV},
pdfsubject={CMS},
pdfkeywords={CMS, physics, disappearing tracks}}

\maketitle

\section{Introduction}
\label{sec:introduction}

This paper presents a search for long-lived, charged particles that decay
within the volume of the silicon tracker of the CMS detector at the CERN LHC
and produce the signature of a ``disappearing track.'' A disappearing track
occurs when the decay products of a charged particle are undetected because
they either have too little momentum to be reconstructed or interact only
weakly, such that they do not produce hits in the tracker and do not deposit
significant energy in the calorimeters.

Anomaly-mediated supersymmetry breaking (AMSB) \cite{Giudice:1998xp,
Randall:1998uk} is one of the many beyond-the-standard-model (BSM) scenarios in
which such a disappearing track would be produced, and one that has been widely
used to interpret the results of searches for disappearing tracks. In AMSB, a
particle mass spectrum is predicted with a small mass difference between the
lightest chargino (\PSGcpmDo) and neutralino (\PSGczDo), where the latter is
the lightest supersymmetric
particle~\cite{Chen:1996ap,Cheng:1998hc,Gherghetta:1999sw,Feng:1999fu}. The
chargino decays to a neutralino and a pion: $\PSGcpmDo \to \PSGczDo \Pgppm$.
Because of the small chargino-neutralino mass difference, the phase space for
this decay is limited, and as a consequence the chargino has a lifetime on the
order of 1\unit{ns}. The pion from this decay has low momentum
($\approx$100\MeV), generally too low for it to be observable as a
reconstructed track. If the chargino decays inside the tracker volume, it thus
will often leave a disappearing track. We present the search in terms of the
chargino mass and lifetime in AMSB, although other BSM scenarios that produce a
disappearing track signature have also been proposed~\cite{Chen:1996ap,
Ibe:2012hu, Hall:2012zp, Arvanitaki:2012ps, ArkaniHamed:2012gw, Citron:2012fg}.

Previous analyses performed by the CMS and ATLAS Collaborations have searched
for disappearing tracks in proton-proton ($\Pp\Pp$) collision data at $\sqrt{s}
= 8\TeV$~\cite{CMS:2014gxa,Aad:2013yna}, and a recent analysis by the ATLAS
Collaboration searched for short disappearing tracks in 13\TeV
data~\cite{Aaboud:2017mpt}. The previous CMS search excluded at 95\% confidence
level (\CL) direct electroweak production of charginos with a mass less than
505\GeV for a mean proper lifetime of 7\unit{ns}, while the ATLAS search at
13\TeV extended the exclusion limits on chargino mass to 460\GeV for a lifetime
of 0.2\unit{ns}. These searches are complementary to searches for heavy stable
charged particles, which are able to exclude charginos with much longer
lifetimes~\cite{Chatrchyan:2013oca,Khachatryan:2015lla}. Two significant
improvements with respect to the 8\TeV search for disappearing tracks have been
implemented for this search at 13\TeV: a new dedicated trigger developed
specifically for this search, and an estimation of the background from standard
model (SM) leptons entirely based on control samples in data.

\section{The CMS detector}

The central feature of the CMS apparatus is a superconducting solenoid of
6\unit{m} internal diameter. Within the solenoid volume are a silicon pixel and
strip tracker, a lead tungstate crystal electromagnetic calorimeter (ECAL), and
a brass and scintillator hadron calorimeter (HCAL), each composed of a barrel
and two endcap sections. Forward calorimeters extend the pseudorapidity
coverage provided by the barrel and endcap detectors. Muons are measured in
gas-ionization detectors embedded in the steel flux-return yoke outside the
solenoid.

The silicon tracker measures charged particles within the pseudorapidity range
$\abs{\eta} < 2.5$. It consists of 1440 silicon pixel and 15\,148 silicon strip
detector modules and is located in the 3.8\unit{T} field of the solenoid.  For
particles that are not explicitly required to be isolated from other event
activity, and that have transverse momentum $1 < \pt < 10\GeV$ and $\abs{\eta}
< 1.4$, the track resolutions are typically 1.5\% in \pt and 25--90
(45--150)\mum in the transverse (longitudinal) impact parameter. These
particles represent the bulk of those produced in collisions. For comparison,
isolated particles of $\pt = 100\GeV$ emitted at $\abs{\eta} < 1.4$ have track
resolutions of 2.8\% in \pt and 10 (30)\mum in the transverse (longitudinal)
impact parameter~\cite{Chatrchyan:2014fea}.

Events of interest are selected using a two-tier trigger system
\cite{Khachatryan:2016bia}. The first level (L1), composed of custom hardware
processors, uses information from the calorimeters and muon detectors to select
events at a rate of around 100\unit{kHz} within a time interval of less than
4\mus. The second level, known as the high-level trigger (HLT), consists of a
farm of processors running a version of the full event reconstruction software
optimized for fast processing, and reduces the event rate to around 1\unit{kHz}
before data storage.

A more detailed description of the CMS detector, together with a definition of
the coordinate system used and the relevant kinematic variables, can be found
in Ref.~\cite{Chatrchyan:2008aa}.

\section{Data sets}

This search uses $\Pp\Pp$ collision data corresponding to an integrated
luminosity of 38.4\fbinv~\cite{CMS:2016eto, CMS:2017sdi}, collected with the
CMS detector at $\sqrt{s} = 13\TeV$ during 2015 and 2016. We analyze separately
the data collected during each of the two years. Further, because of changes to
the trigger configuration during the 2016 run, we also consider the earlier and
later data-taking periods, designated as 2016A and 2016B, separately. The three
running periods, which we analyze independently, and the corresponding
integrated luminosities are presented in Table~\ref{tab:datasets}.

\begin{table}[htbp]
  \centering
  \topcaption{The data-taking periods and the corresponding integrated
  luminosities.}
  \begin{tabular}{lc}
  \hline \\[-2.3ex]
  Run period & Integrated luminosity [$\mathrm{fb}^{-1}$] \\[0.2ex]
  \hline
  2015  & 2.7  \\
  2016A & 8.3  \\
  2016B & 27.4 \\
  \hline
  \end{tabular}
  \label{tab:datasets}
\end{table}

Simulated signal events of $\Pp\Pp \to \PSGcpmDo \PSGcmpDo$ and $\Pp\Pp \to
\PSGczDo\PSGcpmDo$ are generated at leading order (LO) precision with \PYTHIA
6.4.26 \cite{Sjostrand:2006za} with the CTEQ6L1~\cite{Pumplin:2002vw} parton
distribution function (PDF) set for \PSGcpmDo masses from 100 to 900\GeV and
lifetimes from 0.33\unit{ns} to 330\unit{ns}, using sparticle mass spectra
produced by \ISAJET 7.80 \cite{Paige:2003mg}. The branching fraction for
$\PSGcpmDo \to \PSGczDo \Pgppm$ is set to 100\%, and \tanb is fixed to 5 with
$\mu > 0$, where \tanb is the ratio of the vacuum expectation values of the two
Higgs doublets and $\mu$ is the higgsino mass parameter. In practice the
\PSGcpmDo-\PSGczDo mass difference has little dependence on \tanb and the sign
of $\mu$~\cite{Ibe:2012sx}. These events are normalized using chargino
production cross sections calculated at next-to-leading order (NLO) plus
next-to-leading-logarithmic (NLL) precision using \textsc{Resummino}
1.0.9~\cite{Fuks:2012qx,Fuks:2013vua} with CTEQ6.6~\cite{Nadolsky:2008zw} and
MSTW2008nlo90cl~\cite{Martin:2009iq} PDF sets. The final cross sections and
uncertainties are calculated using the PDF4LHC recommendations for the two sets
of cross sections~\cite{Butterworth:2015oua}. The ratio of $\PSGczDo\PSGcpmDo$
to $\PSGcpmDo\PSGcmpDo$ production is estimated to be roughly 2:1 for all
chargino masses considered. Scale factors are applied as a function of the \pt
of the sparticle pair (either $\PSGcpmDo\PSGcmpDo$ or $\PSGczDo\PSGcpmDo$) to
correct for mismodeling of initial state radiation (ISR) in \PYTHIA; they are
derived by comparing experimental and simulated data in a control region
populated mainly by \Ztomumu decays as a function of the \pt of the \cPZ\ boson
candidate, similar to the method used in Ref.~\cite{Chatrchyan:2013xna}. These
events were chosen because the production modes of the \cPZ\ boson and the
\PSGcpmDo are similar. The scale factors typically result in a correction of
order $+$25\% in the kinematic region relevant to this search.

Although the methods used to estimate backgrounds in this search are based on
experimental data, samples of simulated SM processes are used to validate them
and calculate systematic uncertainties. Drell--Yan events, single top quark
production via the $s$ and $t$ channels, $\cPZ\gamma$, $\PW\gamma$, and \Wtolnu
events, where $\ell$ can be an electron, muon, or tau lepton, are generated at
NLO precision using the \MGvATNLO~2.3.3 generator~\cite{aMCNLO}. The $\PW\cPZ$,
$\cPZ\cPZ$, and quantum chromodynamics (QCD) multijet events, with the last
composed of jets produced solely through the strong interaction, are generated
at LO precision with \PYTHIA~8.205~\cite{Sjostrand:2007gs}. The $\PW\PW$,
\ttbar, $\cPqt\PW$, and $\cPaqt\PW$ events are generated at NLO precision using
\POWHEG~v2.0~\cite{Nason:2004rx,Frixione:2007vw,Alioli:2010xd,Melia:2011tj,Nason:2013ydw,Frixione:2007nw,Re:2010bp}.
The fragmentation and hadronization for all simulated background processes are
provided by \PYTHIA~8.205. The NNPDF3.0~\cite{Ball:2014uwa} PDF set is used for
all simulated backgrounds, and the CUETP8M1~\cite{Monash,CUETP8M1} tune is used
for the underlying event.

For both simulated signal and background events, the detector response is
described by a full model of the CMS detector based on
\GEANTfour~\cite{Agostinelli:2002hh} and reconstructed with the same software
that is used for collision data. Simulated minimum bias events are superimposed
on the hard interaction to describe the effect of overlapping inelastic
$\Pp\Pp$ interactions within the same or neighboring bunch crossings, known as
pileup, and the samples are reweighted to match the reconstructed vertex
multiplicity observed in data.

\section{Event reconstruction and selection}

The particle-flow (PF) event algorithm~\cite{CMS-PRF-14-001} is designed to
reconstruct and identify each individual particle with an optimized combination
of information from the various elements of the CMS detector. The energy of
photons is directly obtained from the ECAL measurement, corrected for
zero-suppression effects. The energy of electrons is determined from a
combination of the electron momentum at the primary interaction vertex as
determined by the tracker, the energy of the corresponding ECAL cluster, and
the energy sum of all bremsstrahlung photons spatially compatible with
originating from the electron track. The energy of muons is obtained from the
curvature of the corresponding track. The energy of charged hadrons is
determined from a combination of their momentum measured in the tracker and the
matching ECAL and HCAL energy deposits, corrected for zero-suppression effects
and for the response function of the calorimeters to hadronic showers. Finally,
the energy of neutral hadrons is obtained from the corresponding corrected ECAL
and HCAL energy.

This search is performed on events that pass one or more of several triggers
with requirements on missing transverse momentum, a characteristic of signal
events where the missing transverse momentum is generated by an ISR jet
recoiling off the sparticle pair. For this specific analysis we define the
vector \ptvecmiss, with magnitude \ptmiss, as the projection onto the plane
perpendicular to the beam axis of the negative vector sum of the momenta of all
reconstructed PF candidates in an event, with the exception of muons, or, in
the case of the L1 trigger, of all calorimeter energy deposits. The triggers
require \ptmiss at L1, with the specific requirement varying throughout data
taking with changes in the instantaneous luminosity. At the HLT, events with
either \ptmiss or \metWithMu, which is defined similarly to \ptmiss but with
muons included in its calculation, are selected. The lowest-threshold trigger,
which was developed specifically for this search, requires $\ptmiss > 75\GeV$
as well as an isolated track with $\pt > 50\GeV$ at the HLT. The
higher-threshold triggers require either \ptmiss or \metWithMu to be greater
than 90 (120)\GeV for the 2015 (2016) data. For signal events, which typically
have no reconstructed muons, \ptmiss and \metWithMu are usually identical, and
both are used at the HLT to mitigate any inefficiency in the isolated track
requirement for events with higher \ptmiss or \metWithMu. In the offline
selection, only \ptmiss is used, in order to mirror the requirements in the L1
trigger and lowest-threshold HLT path.  Events are required to have $\ptmiss >
100\GeV$ offline, where $\ptmiss$ is calculated from the full PF
reconstruction.

Jets are clustered from PF candidates using
\FASTJET~3.10~\cite{Cacciari:2011ma} with the anti-\kt
algorithm~\cite{Cacciari:2008gp} with a distance parameter of 0.4, and only
jets with $\pt > 30\GeV$ and $\abs{\eta} < 2.4$ are considered in the analysis.
Additional criteria are imposed on these jets to remove those originating from
calorimeter noise and misidentified leptons~\cite{CMS-PAS-JME-16-003}. Events
are required to have at least one jet with $\pt > 110\GeV$ in order to be
consistent with the ISR recoil topology.

\begin{figure}[htbp]
  \centering
  \includegraphics[width=0.49\textwidth]{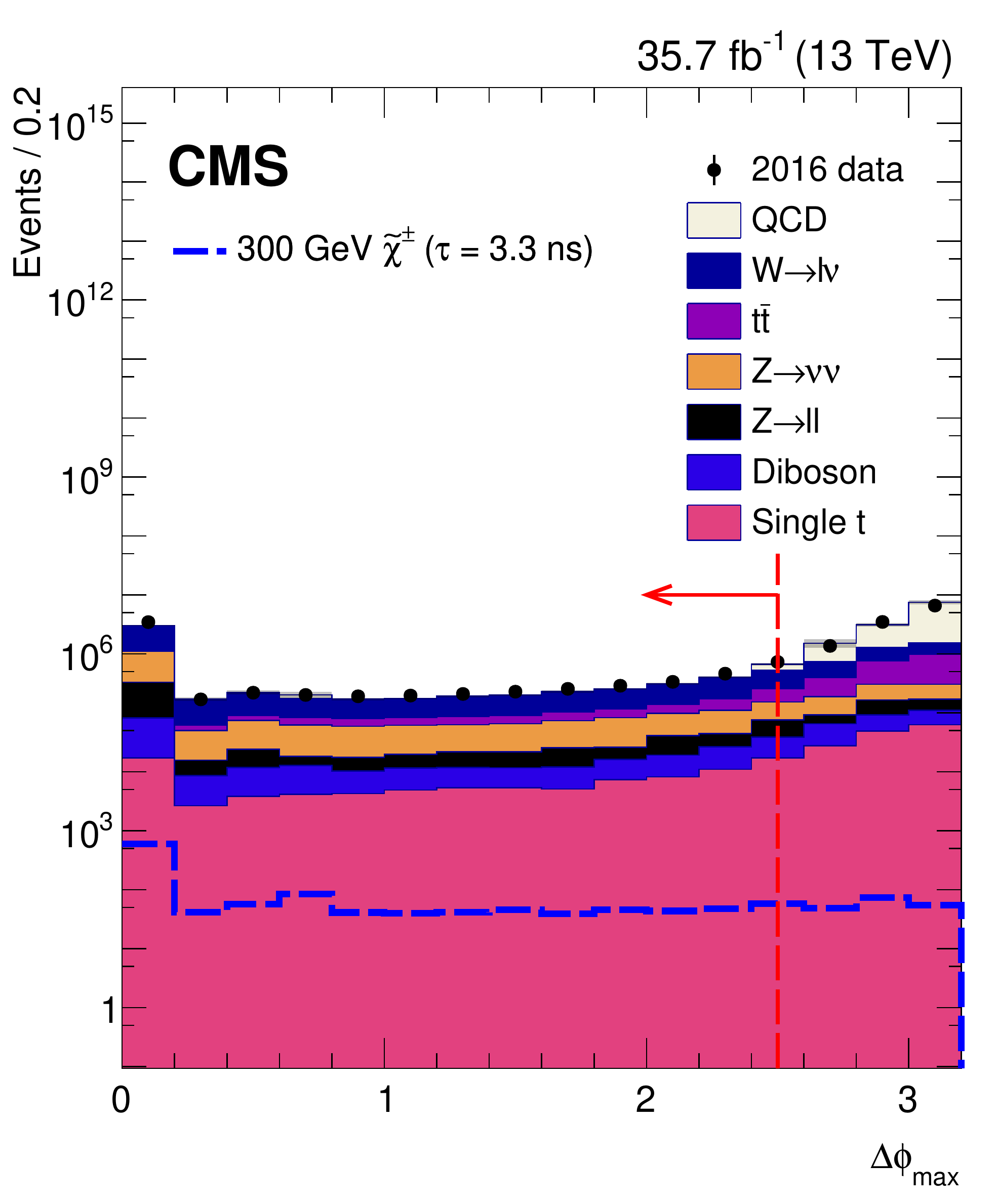}
  \includegraphics[width=0.49\textwidth]{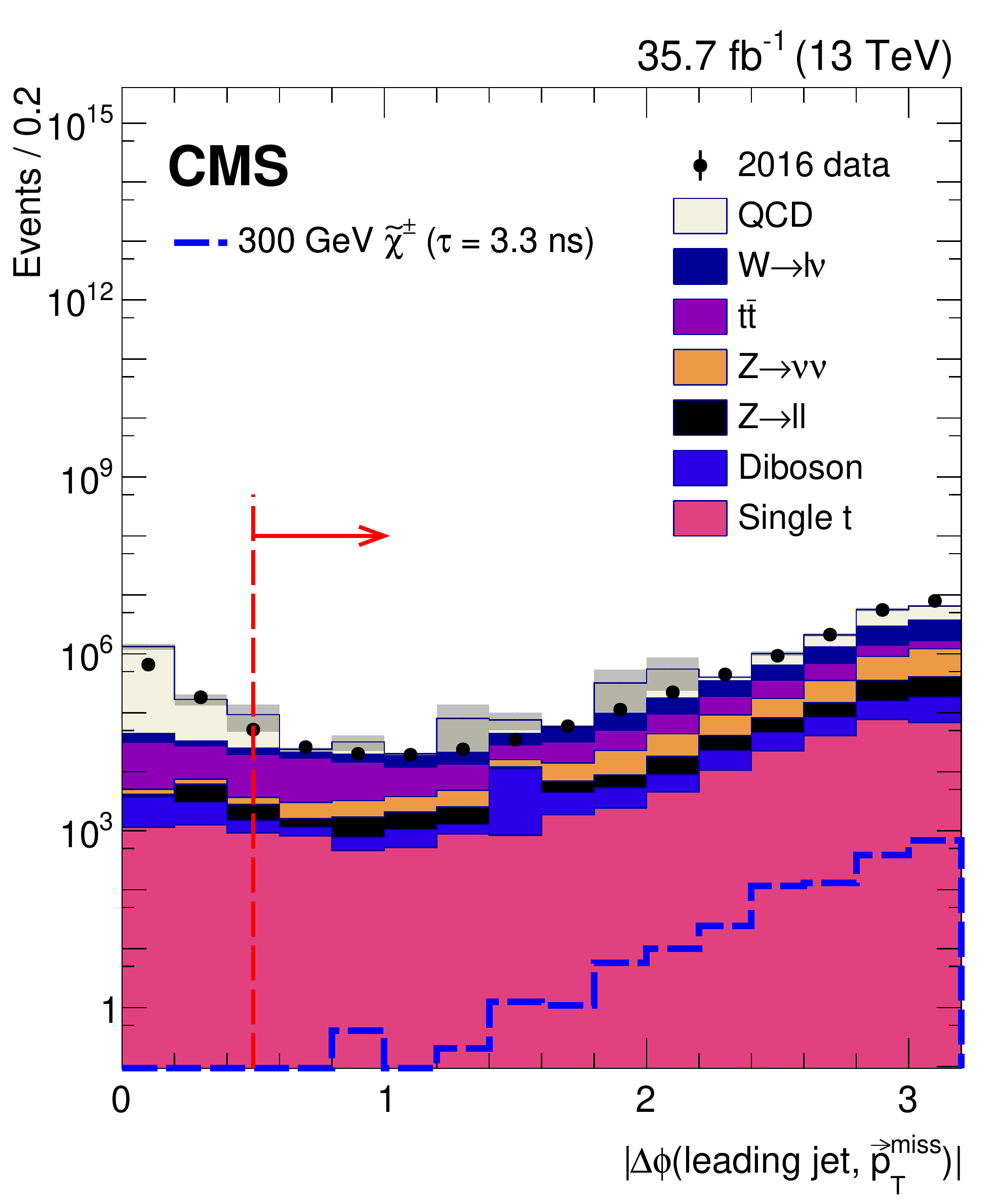}
  \caption{Distributions of the maximum difference in $\phi$ between any two
  jets (\cmsLeft) and the difference in $\phi$ between the \ptvec of the
  leading jet and \ptvecmiss (\cmsRight) for events passing the basic
  selection, before either of the requirements on these two variables is
  imposed. The data is from the 2016 data-taking period, and the blue dashed
  lines show the distributions for simulated signal events with a chargino that
  has a lifetime of 3.3\unit{ns} and mass of 300\GeV, with a corresponding
  production cross section of 0.58\unit{pb}. The gray shaded area indicates the
  statistical uncertainty in the SM background, and the leftmost bin of the
  left plot includes events with only one selected jet. The vertical dashed
  lines indicate the chosen value for the requirement on each variable, and the
  arrows indicate which events are selected.}
  \label{fig:deltaPhi}
\end{figure}

We require the difference in azimuthal angle, $\phi$, between the \ptvec of the
leading (highest energy) jet and \ptvecmiss to be greater than 0.5, and for
events with at least two jets, we require the maximum difference in $\phi$
between any two jets, $\Delta \phi_{\text{max}}$, to be less than 2.5. These
requirements are designed to remove the large, reducible background originating
from QCD multijet events. In these events, a dijet topology with back-to-back
jets dominates and mismeasurement of the jet energy may result in a significant
measured $\ptmiss$. We refer to the selection up to this point, before any
track-related criteria are imposed, as the ``basic selection.'' Events passing
this selection are expected to have minimal signal contamination and are
dominated by the \Wtolnu process. The effect of the two angular requirements of
the basic selection on the 2016 data and on simulated signal and background
events is shown in Fig.~\ref{fig:deltaPhi}. For signal events, the shapes of
these distributions are largely independent of chargino mass and lifetime, and
a single representative signal point is shown. The combination of these two
requirements is sufficient to remove most of the large QCD multijet background
that would otherwise pass the basic selection, while the majority of the
remaining background is removed by the track criteria described below.

After the basic selection, tracks are selected that have $\pt > 55\GeV$ and
$\abs{\eta} < 2.1$. The track \pt requirement is chosen such that the
corresponding requirement in the HLT path is fully efficient. We ensure that
selected tracks are isolated from other activity in the tracker by requiring
the scalar sum of the \pt of other tracks within a cone of $\DR =
\sqrt{\smash[b]{(\Delta\phi)^2 + (\Delta\eta)^2}} < 0.3$ around the momentum
vector of the selected track be less than 5\% of the \pt of the track. Selected
tracks are also required to be well-separated from jets with $\DR
(\text{track},\,\text{jet}) > 0.5$.

One source of background for this search arises from ``spurious tracks,'' \ie,
pattern recognition errors that do not correspond to actual charged particles.
Spurious tracks can have missing hits in the outer layers of the silicon
tracker and muon detectors, and are not generally associated with large energy
deposits in the calorimeters, thus mimicking a disappearing track. This
background is suppressed by requiring that selected tracks have at least three
hits in the pixel detector and at least seven hits overall in the tracker, a
typical non-disappearing track leaving twice that number of hits on average. A
missing hit in a layer of the tracker between the interaction point and the
first actual hit on the track is called a missing inner hit, while a missing
hit between the first and last hits on the track is called a missing middle
hit. We require selected tracks to have no missing inner or middle hits. In
other words, there must be a consecutive pattern of hits originating in the
tracker layers closest to the interaction point. Since spurious tracks often
appear displaced from the interaction point, we also require all tracks to have
a transverse impact parameter $\abs{\dz} < 0.02\cm$ and a longitudinal impact
parameter $\abs{z_{0}} < 0.5\cm$, both with respect to the primary vertex,
chosen as the reconstructed vertex with the largest value of summed physics
object $\pt^2$. The physics objects are the jets, clustered using the jet
finding algorithm~\cite{Cacciari:2008gp,Cacciari:2011ma} with the tracks
assigned to the vertex as inputs, and the associated missing transverse
momentum, taken as the negative vector sum of the \pt of those jets. More
details are given in Section~9.4.1 of Ref.~\cite{CMS-TDR-15-02}.

Besides spurious tracks, most isolated, high-\pt tracks from SM processes come
from charged leptons produced in the decays of \PW\ or \cPZ\ bosons or virtual
photons. Thus, the other main source of background for this search arises from
isolated charged leptons that are not correctly reconstructed by the PF
algorithm. Leptons can have missing hits in the tracker for several reasons:
for example, energetic bremsstrahlung in the case of electrons, or nuclear
interactions with the tracker material in the case of hadronically decaying tau
leptons (\tauh). Leptons may also have small associated calorimeter energy
deposits because of nonoperational or noisy channels. To mitigate this
background, events where selected tracks are close to reconstructed leptons
($\DR (\text{track},\,\text{lepton}) < 0.15$) are vetoed. To avoid selecting
leptons that fail to be reconstructed because of detector inefficiencies, we
impose the following fiducial track criteria. We avoid regions of muon
reconstruction inefficiency by vetoing tracks within gaps in the coverage of
the muon chambers at $0.15 < \abs{\eta} < 0.35$ and $1.55 < \abs{\eta} < 1.85$.
Similarly, we avoid regions of electron reconstruction inefficiency by
rejecting tracks within the overlap region between the barrel and endcap
sections of the ECAL at $1.42 < \abs{\eta} < 1.65$, as well as tracks within
$\DR < 0.05$ of a nonoperational or noisy ECAL channel, where \DR\ is
calculated with respect to the track at the point of closest approach to the
center of CMS.

Additional areas of inefficiency are identified using electron and muon
tag-and-probe (T\&P) studies~\cite{Khachatryan:2010xn}, where \Ztoll candidates
are selected in data with $\Mll \approx \MZ$, where \MZ is the world-average
mass of the \cPZ\ boson~\cite{Patrignani:2016xqp}, and the \cPZ\ resonance is
exploited to obtain a sample of tracks that have a high probability of being
leptons, without explicitly requiring them to be reconstructed as leptons. The
fraction of these tracks that are not explicitly identified as leptons passing
a loose set of identification criteria is a measure of the inefficiency for
identifying leptons and is grouped in bins in the $\eta$-$\phi$ plane. Tracks
in bins with an anomalously high inefficiency are rejected from the selection.
This procedure removes $\approx$4\% of tracks in simulated signal events that
would otherwise be selected.

\begin{figure}[htbp]
  \centering
  \includegraphics[width=0.49\textwidth]{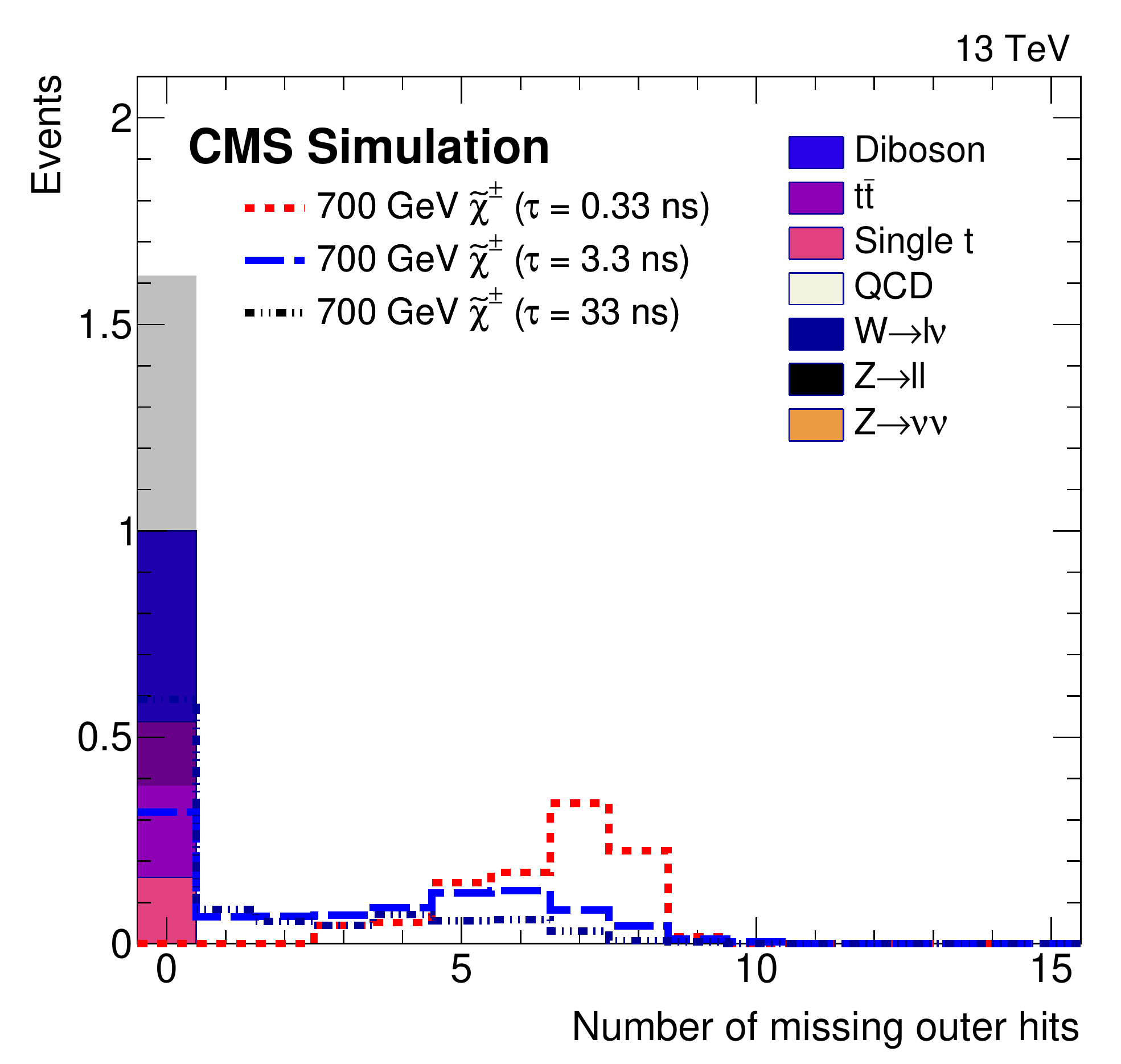}
  \caption{Distributions of the number of missing outer hits for tracks in
  simulation that pass the full selection, except for the requirement on that
  variable. Each signal distribution and the sum of the SM background
  distributions are scaled to have unit area. The gray shaded area indicates
  the statistical uncertainty in the SM background.}
  \label{fig:nMinusOne}
\end{figure}

Two additional requirements define the criteria for a track to ``disappear.''
First, we require the selected tracks to have at least three missing outer
hits, which are missing hits in the tracker layers outside of the last layer
containing a hit on the track. Second, the associated calorimeter energy within
$\DR < 0.5$ of the track, \Ecalo, is required to be less than 10\GeV, where
\DR\ is calculated using the track coordinates at the point of closest approach
to the center of CMS. This requirement removes a negligible amount of signal,
while \Ecalo is much larger, typically over 100\GeV, for background events
passing the other selection criteria, according to the simulation. The number
of missing outer hits is shown in Fig.~\ref{fig:nMinusOne} for simulated signal
and background events that pass the full selection, except for the requirement
on that variable. The tracks selected in the simulated background events are
predominantly from electrons and \tauh, since events with muons have a smaller
$\ptmiss$ on average. As can be seen, the number of missing outer hits is very
effective at isolating the signal because tracks from background events
typically have no missing outer hits. The efficiency of the full selection for
simulated signal events is limited mostly by the requirements targeting events
with ISR and the relatively narrow range of chargino decay lengths that yield a
disappearing track that passes the criterion on the number of missing outer
hits. This efficiency varies with the chargino mass and lifetime, peaking at
$\approx$2\% for a 700\GeV chargino with a lifetime of 3\unit{ns}.

\section{Background estimation}
\label{sec:backgroundEstimation}

\subsection{Charged leptons}

The dominant source of high-\pt, isolated tracks from SM processes arises from
charged leptons (electrons and muons, produced promptly or via the decay of tau
leptons, or \tauh) from the decay of \PW\ or \cPZ\ bosons or virtual photons.
In order for events with such tracks to appear in the search region, three
things must happen: (1) the lepton must fail to be explicitly identified as a
lepton, while still leaving a track in the silicon tracker but less than 10\GeV
of energy in the calorimeters; (2) the resulting \metWithMu and \ptmiss must be
large enough for the event to pass the triggers; and (3) the resulting \ptmiss
must be large enough for the event to pass the offline \ptmiss requirements.

The key point is that \metWithMu and \ptmiss are affected by whether the lepton
is explicitly identified as a lepton or not. If it is not, but still leaves a
track in the silicon tracker and less than 10\GeV of energy in the
calorimeters, its energy does not typically contribute to the visible energy of
the event. The method used to estimate the background from charged leptons is
based on calculating the probability in data of the three conditions listed
above, with each lepton flavor treated independently.

The first probability we consider is \Pveto, the probability that the lepton in
a single-lepton event is not explicitly identified as a lepton. For each flavor
of charged lepton, we estimate \Pveto using a T\&P study. The electron (muon)
T\&P selections utilize \Ztoee ($\Pgm\Pgm$) candidates. For this study we
select events passing a single-electron (single-muon) trigger and containing a
reconstructed electron (muon) that passes tight identification and isolation
criteria. This lepton serves as the tag. A probe track is required to pass the
disappearing track criteria, except for those defining the electron (muon) veto
in Table~\ref{tab:leptonVetoes}. The tag lepton and the probe track are
required to have an invariant mass within 10\GeV of the \cPZ\ boson mass and to
have opposite signs of electric charge.

\begin{table}[htbp]
  \centering
  \topcaption{Definitions of the lepton vetoes used in the T\&P studies to
  estimate \Pveto, for each flavor of charged lepton. The criteria listed are
  the subset of the search criteria that are the most efficient at rejecting
  each flavor.}
  \begin{tabular}{lccc}
  \hline
  Selection & Electron & Muon & Tau lepton \\
  \hline
  $\operatorname{Min} \DR_{\text{track},\,\text{electron}} > 0.15$ & \checkmark & & \\
  $\operatorname{Min} \DR_{\text{track},\,\text{muon}} > 0.15$ & & \checkmark & \\
  $\operatorname{Min} \DR_{\text{track},\,\tauh} > 0.15$ & & & \checkmark \\
  $\Ecalo < 10\GeV$ & \checkmark & & \checkmark \\
  Missing outer hits $\geq 3$ & \checkmark & \checkmark & \checkmark \\
  $\DR_{\text{track},\,\text{jet}} > 0.5$ & & & \checkmark \\[0.3ex]
  \hline
  \end{tabular}
  \label{tab:leptonVetoes}
\end{table}

For the \tauh T\&P study, we define two selections using \Ztotautau events that
are combined for the calculation of \Pveto: one where the electron from a $\Pgt
\to \Pe\nu\nu$ candidate is selected as the tag, and one where the muon from a
$\Pgt \to \Pgm\nu\nu$ candidate is selected as the tag. These two selections
are identical to the electron and muon T\&P selections defined above,
respectively, except for two modifications. First, we require the transverse
mass $\MT = \sqrt{\smash[b]{2 \pt^\ell \metWithMu (1 - \cos \Delta \phi)}}$ to
be less than 40\GeV, where $\pt^\ell$ is the magnitude of the transverse
momentum of the tag lepton and $\Delta \phi$ is the difference in $\phi$
between the \ptvec of the tag lepton and \metWithMuVec. This \MT requirement is
made to reduce contamination from \Wtolnu events. Second, because the \Pgt\
leptons from the \cPZ\ decay are not fully reconstructed, the dilepton
invariant mass requirement is $\MZ - 50 < m < \MZ - 15\GeV$.

For each of these selections, we also define a version in which the tag lepton
and the probe track are required to have the same sign of electric charge
instead of opposite signs. This requirement makes it unlikely that the selected
probe track candidates are genuine tracks, and these selections are used to
subtract the background from spurious tracks in the calculation of \Pveto.

For each of the three T\&P channels (electrons, muons, and \tauh), the
quantities \Ntp (\NtpVeto) and \NSStp (\NSStpVeto) are the numbers of selected
T\&P pairs before (after) the final lepton veto is applied to the probe tracks,
for the opposite-sign and same-sign selection, respectively. From this, the
veto probability is calculated as:
\begin{linenomath} \begin{equation}
  \Pveto = \frac{\NtpVeto - \NSStpVeto}{\Ntp - \NSStp}.
\end{equation} \end{linenomath}

We define \Poffline as the conditional probability of a single-lepton event to
pass the offline requirements of $\ptmiss > 100\GeV$ and $\dPhiJetMET > 0.5$
given that the lepton candidate is not explicitly identified as a lepton. Using
events in single-lepton control regions in data, we introduce a modified
\ptvecmiss variable that represents what \ptvecmiss would look like if the
lepton in these events were not explicitly identified as such, assuming that if
a lepton is not explicitly identified it contributes no visible energy to the
event. In the single-electron and single \tauh control regions, we use
\metMinusOneVec. For the single-muon control region, we simply use \ptvecmiss
since the \pt of all reconstructed muons is already excluded from its
calculation. We then estimate \Poffline by counting the fraction of events with
$\ptmiss > 100\GeV$ and $\dPhiJetMET > 0.5$ after modifying \ptvecmiss in this
way.

We define \Ptrigger as the conditional probability of a single-lepton event to
pass the \metWithMu or \ptmiss triggers, given that the lepton candidate is not
explicitly identified as a lepton and that the event passes the offline
requirements of $\ptmiss > 100\GeV$ and $\dPhiJetMET > 0.5$. The estimation of
\Ptrigger is made in a similar way to the estimation of \Poffline in the
single-lepton control regions, assuming that a lepton that is not explicitly
identified as such contributes no visible energy to the event and constructing
the modified \metMinusOneVec for electrons and \tauh, using \ptvecmiss for
muons. The exception for \Ptrigger is that instead of constructing these
quantities with offline reconstructed leptons, online objects are used from
both the L1 trigger and the HLT. For each lepton selected in each of the
single-lepton control regions in data, we find the closest L1 trigger object
and closest HLT object within $\DR < 0.1$ of the offline object. The \ptvec of
these objects is then added to the nominal \ptvecmiss, as calculated by the L1
trigger and HLT, respectively, and to the nominal \metWithMuVec in the case of
the HLT. This way, we can test, event by event, if the L1 trigger and HLT would
have passed, given these modifications to the online \metWithMuVec and
\ptvecmiss. The number of events passing the offline \ptmiss requirements is
calculated following the procedure used to calculate \Poffline, and the
fraction of these events that also pass the \metWithMu and \ptmiss triggers
according to the above procedure is then the estimate of \Ptrigger.

The product of the three probabilities defined above (\Pveto, \Poffline, and
\Ptrigger) gives the probability of an event with a charged lepton to enter the
search region. We use the single-lepton control regions to estimate the total
numbers of events in data containing each flavor of lepton, \Nctrll, and obtain
the estimated number of background events from charged leptons as
\begin{linenomath} \begin{equation}
  \Nestl = \Nctrll \Pveto \Poffline \Ptrigger.
\end{equation} \end{linenomath}

Closure tests were performed with samples of simulated background events and
with the early 13\TeV data taken in 2015. Both tests proved the validity of the
background estimation method, with agreement within $1.2\,\sigma\xspace$
observed in all cases.

\subsection{Spurious tracks}

The contribution of spurious tracks to the background is largely suppressed by
the requirement that the impact parameters of the tracks with respect to the
primary vertex are small and by the requirement that the tracks are missing no
inner or middle hits in the tracker. We estimate the residual contribution from
this background using a control region of \Ztomumu events as a representative
sample of SM events. Within this sample, we additionally require a track,
separate from the muons coming from the \cPZ\ boson candidate, that passes the
track requirements of the search region except for the transverse impact
parameter criterion, which we replace with a sideband selection, $0.02 <
\abs{\dz} < 0.10\cm$, designed to enhance the likelihood that the tracks we
select are spurious. In this way, we can estimate the probability for there to
be spurious tracks that satisfy these requirements. This probability is
multiplied by a transfer factor to obtain the probability of spurious tracks
passing the nominal impact parameter requirement, \Pfake. This transfer factor
is obtained from a sample of tracks with only three hits in the pixel detector
and no hits in the strip detector, which is dominated by spurious tracks. The
estimated background from spurious tracks is the number of events in data
passing the basic selection, \NctrlBasic, multiplied by \Pfake:
\begin{linenomath} \begin{equation}
  \NestFake = \NctrlBasic \Pfake.
\end{equation} \end{linenomath}

\section{Systematic uncertainties}
\label{sec:systematics}

\subsection{Background estimates}
\label{sec:backgroundSystematics}

The lepton background estimates rely on the assumption that when a lepton is
not explicitly identified as a lepton, while still leaving a track in the
silicon tracker but less than 10\GeV of energy in the calorimeters, it
contributes no visible energy to the event. We test the impact of this
assumption for electrons and $\tauh$ by replacing the nominal \metMinusOneVec
variable used to calculate \Poffline and \Ptrigger with a ``scaled down''
version,
\begin{linenomath} \begin{equation}
  \metMinusOneUpVec,
\end{equation} \end{linenomath}
\noindent and recalculating \Poffline and \Ptrigger. In other words, we assume
that unreconstructed leptons contribute 10\GeV of visible energy to the event.
The value of 10\GeV is chosen because selected tracks are required to have
$\Ecalo < 10\GeV$ in the disappearing track search region. The difference from
unity of the ratio
\begin{linenomath} \begin{equation}
  \frac{(\Poffline \Ptrigger)_{\text{scaled\,down}}}{(\Poffline \Ptrigger)_{\text{nominal}}}
\end{equation} \end{linenomath}
\noindent is taken as the systematic uncertainty. This uncertainty is
approximately 12 (17)\% for electrons (\tauh) and is not calculated for muons,
since even successfully reconstructed muons are not expected to contribute
substantial visible calorimeter energy to an event.

For the spurious track background estimate, it is assumed that the particular
choice of the \dz sideband region results in predominantly spurious tracks
being selected. To test the impact of this assumption, we examine the
variations in the background estimate as the lower bound on the sideband is
increased from 0.02 to 0.10\cm. These variations are indeed consistent with the
nominal estimate within statistical uncertainties, with maximum variations of
100\% down and 45\% up for the 2016 data, which are assigned as systematic
uncertainties. For the 2015 data, since the estimate is zero and there is no
indication of behavior different from 2016 data, we assign a systematic
uncertainty of 50\% for this data. To apply systematic uncertainties to
estimates of zero events, the recommendations of Ref.~\cite{Cousins:1991qz} are
followed.

A systematic uncertainty associated with the evaluation of the sideband
transfer factor using tracks with three hits is determined. This systematic
uncertainty is evaluated by examining the variation in the \dz distribution
from tracks with three consecutive hits to at least seven consecutive hits
using tracks in simulated events that are not associated with a generated
particle. In this way, we can see how much the true distribution of \dz for
spurious tracks varies with the number of hits, and constrain the impact this
variation has on the background estimate. This procedure yields an uncertainty
of approximately $-$50\% and $+$100\% in the spurious-track background
estimate.

The spurious-track background estimate rests on the assumption that the
spurious-track probability is similar for events in the \Ztomumu control region
and events passing the basic selection. However, there is nothing about the
method used to calculate this probability that prevents us from calculating it
for events passing the basic selection, and we are able to compare the
estimates we obtain from these two independent control regions. This comparison
serves to validate the method for estimating the spurious-track background, and
the relative difference between the estimates is assigned as a systematic
uncertainty. Excellent agreement is seen between the two control regions in
both the spurious-track probability and the spurious-track estimate itself,
with the estimates agreeing to within $\approx$8\% for the 2016 data, and this
is taken as a systematic uncertainty. Again, both estimates are zero in the
2015 data, but without any indication that their behaviors are different from
2016 data, we assign a 20\% systematic uncertainty for this period and
implement this as in Ref.~\cite{Cousins:1991qz}.

\subsection{Signal efficiencies}
\label{sec:signalSystematics}

Theoretical uncertainties of 3--9\% (depending on the chargino mass), which
include factorization and renormalization scale uncertainties as well as the
PDF uncertainties, are assigned to the chargino production cross sections.
Additional sources of systematic uncertainty in the signal yields include those
in the integrated luminosity, 2.3 (2.5)\% for 2015 (2016)
data~\cite{CMS:2016eto, CMS:2017sdi}, and those related to the modeling of
pileup (2--3\%), ISR (8--9\%), jet energy scale and resolution (2--6\%), and
\ptmiss (0.4\%), with the values of these uncertainties depending on chargino
mass and lifetime. We also estimate uncertainties in the efficiency of the
selection criteria on missing inner, middle, and outer hits (1--3, 0.3--3,
and 0--3\%, respectively), and \Ecalo (0.6--1\%), with values that depend on
the run period being considered. We evaluate uncertainties to account for
potential mismodeling of the trigger efficiency (4--6\%, depending on chargino
mass and lifetime) and track reconstruction efficiency, namely, 1.5 (4.5)\% for
2015 (2016) data. The systematic uncertainties in the signal yields are
summarized in Table~\ref{tab:signalSystematics}.

\begin{table}[htbp]
  \centering
  \topcaption{Summary of the systematic uncertainties in the signal yields. The
  ranges represent either the variation with chargino mass and lifetime or with
  the data-taking period used to calculate the uncertainty, depending on the
  source of each uncertainty as described in the text.}
  \begin{tabular}{lr@{}l}
  \hline
  Source of uncertainty            &  \multicolumn{2}{c}{Range\,[\%]} \\
  \hline
  Theory                           &  3--&9 \\
  Integrated luminosity            &  2.3--&2.5 \\
  Pileup                           &  2--&3 \\
  ISR                              &  8--&9 \\
  Jet energy scale/resolution      &  2--&6 \\
  \ptmiss modeling               &  \multicolumn{2}{c}{0.4} \\
  Missing inner hits               &  1--&3 \\
  Missing middle hits              &  0.3--&3 \\
  Missing outer hits               &  0--&3 \\
  \Ecalo selection               &  0.6--&1 \\
  Trigger efficiency               &  4--&6 \\
  Track reconstruction efficiency  &  1.5--&4.5 \\[\cmsTabSkip]
  Total                            &  10--&18 \\
  \hline
  \end{tabular}
  \label{tab:signalSystematics}
\end{table}

\section{Results}
\label{sec:results}

The numbers of expected events from background sources compared with the
observation in the search sample are shown in Table~\ref{tab:countSummary}. The
observation agrees with the expected background. We set 95\% \CL upper limits
on the product of the cross section for direct production of charginos
($\sigma$) and their branching fraction to $\PSGczDo\Pgppm$ ($\mathcal{B}$) for
various chargino masses and lifetimes.

\begin{table}[htbp]
  \centering
  \topcaption{Summary of numbers of events for the estimated backgrounds and
  the observed data. The uncertainties include those from statistical and
  systematic sources. In categories where the systematic uncertainty is
  negligible, it is not shown.}
  \begin{tabular}{lcccc}
  \hline
  \multirow{2}{*}{Run period} & \multicolumn{3}{c}{Estimated number of background events} & \multirow{2}{*}{Observed events} \\
  & Leptons & Spurious tracks & Total & \\
  \hline \\[-2.3ex]
  2015 & $0.1 \pm 0.1$ & $0_{-0}^{+0.1}$ & $0.1 \pm 0.1$ & 1 \\
  2016A & $2.0 \pm 0.4 \pm 0.1$ & $0.4 \pm 0.2 \pm 0.4$ & $2.4 \pm 0.5 \pm 0.4$ & 2 \\
  2016B & $3.1 \pm 0.6 \pm 0.2$ & $0.9 \pm 0.4 \pm 0.9$ & $4.0 \pm 0.7 \pm 0.9$ & 4 \\[\cmsTabSkip]
  Total & $5.2 \pm 0.8 \pm 0.3$ & $1.3 \pm 0.4 \pm 1.0$ & $6.5 \pm 0.9 \pm 1.0$ & 7 \\
  \hline
  \end{tabular}
  \label{tab:countSummary}
\end{table}

These limits are calculated using the LHC-type~\cite{CMS-NOTE-2011-005}
modified frequentist \CLs criterion~\cite{Junk:1999kv,Read:2002hq}. This method
uses a test statistic based on a profile likelihood ratio~\cite{Cowan:2010js}
and treats nuisance parameters in a frequentist context. Nuisance parameters
for the theoretical uncertainties in the signal cross sections, and systematic
uncertainties in the integrated luminosity and in the signal selection
efficiency, are constrained with log-normal distributions. There are two types
of nuisance parameters for the uncertainties in the background estimates, and
they are specified separately for each of the four background contributions
(three arising from the three flavors of charged leptons and one from spurious
tracks). Those that result from the limited size of the control samples are
constrained with gamma distributions, while those that are associated with
statistical uncertainties in multiplicative factors and the systematic
uncertainties discussed in Section~\ref{sec:systematics} are constrained with
log-normal distributions.

\begin{figure}[htbp]
  \centering
  \includegraphics[width=0.49\textwidth]{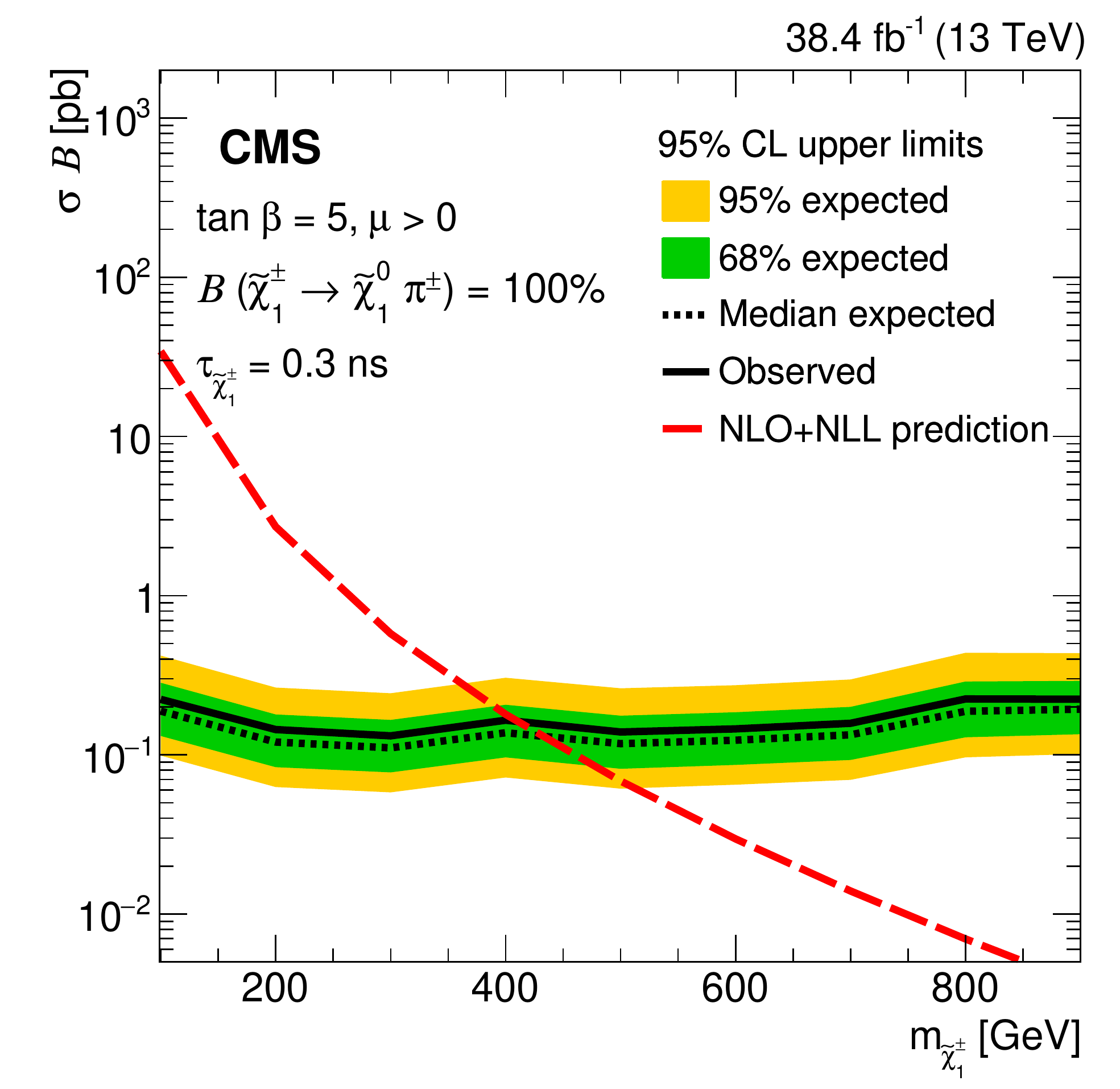}
  \includegraphics[width=0.49\textwidth]{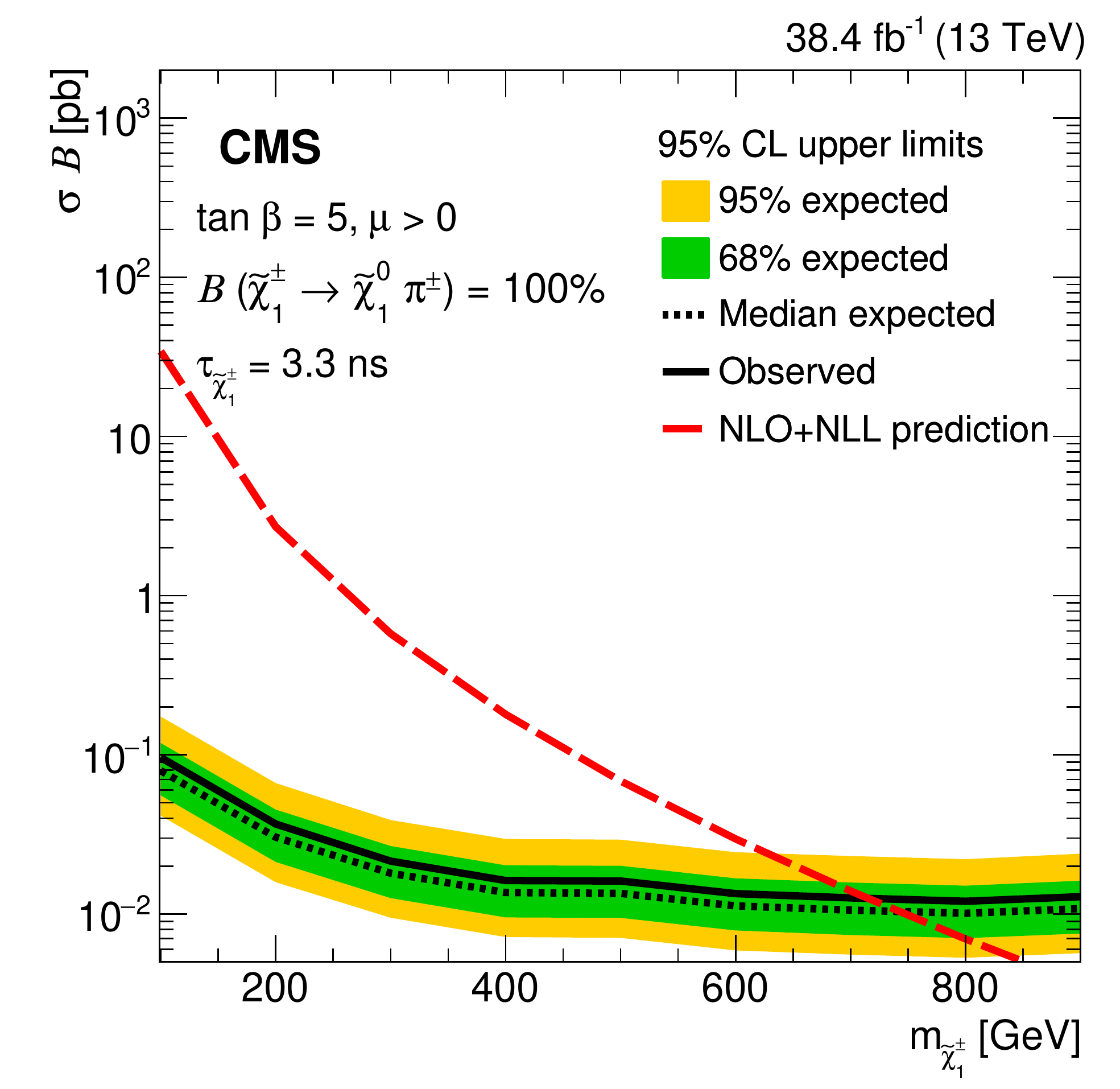} \\
  \includegraphics[width=0.49\textwidth]{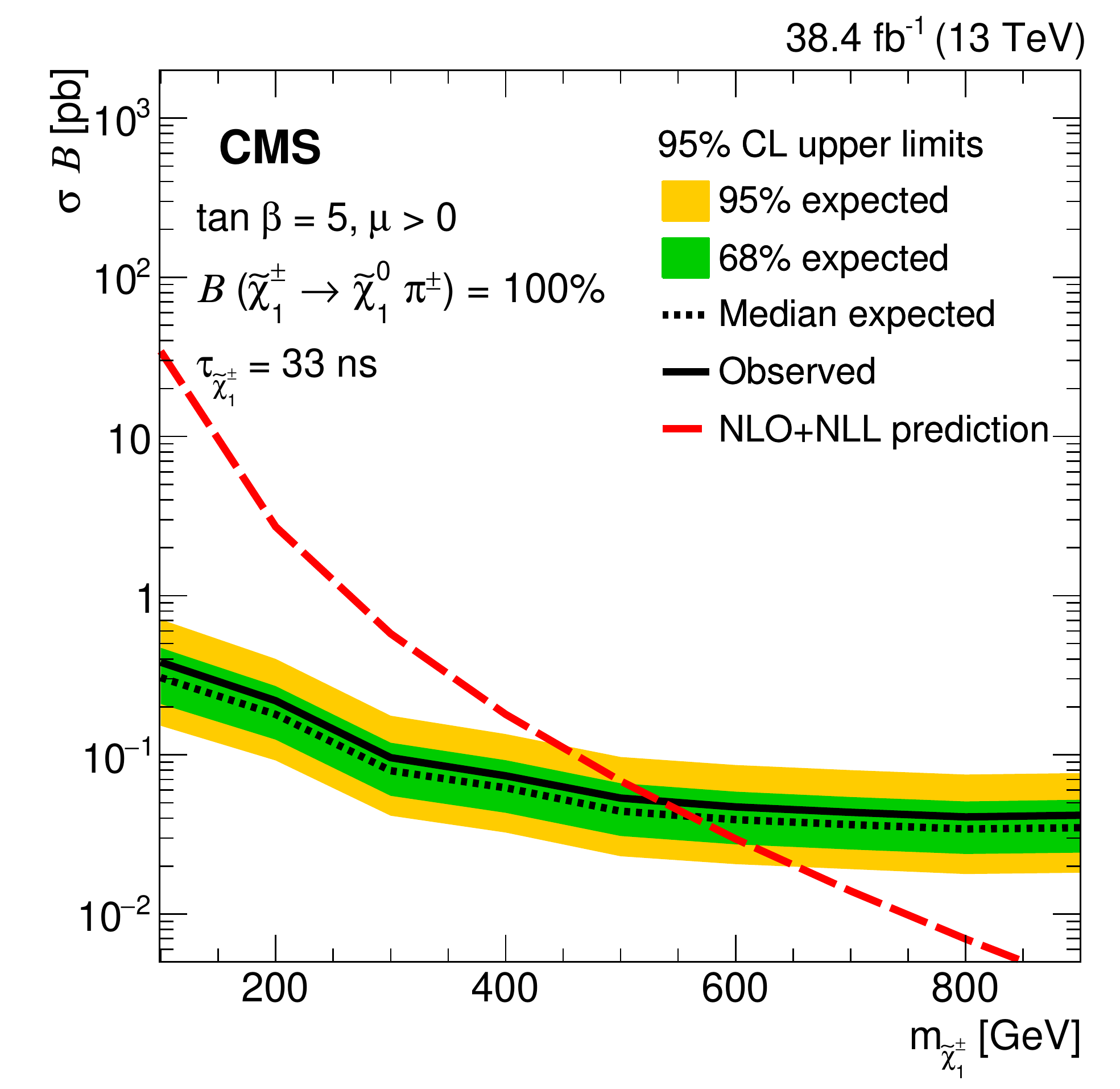}
  \caption{The expected and observed 95\% \CL upper limits on the product of
  the cross section for direct production of charginos and their branching
  fraction to $\PSGczDo\Pgppm$ as a function of chargino mass for chargino
  lifetimes of 0.33, 3.3, and 33 \unit{ns}. The direct chargino production
  cross section includes both $\PSGczDo\PSGcpmDo$ and $\PSGcpmDo\PSGcmpDo$
  production in roughly a 2:1 ratio for all chargino masses considered. The
  dashed red line indicates the theoretical prediction for the AMSB model.}
  \label{fig:limitsCrossSectionVsMass}
\end{figure}

\begin{figure}[htbp]
  \centering
  \includegraphics[width=0.59\textwidth]{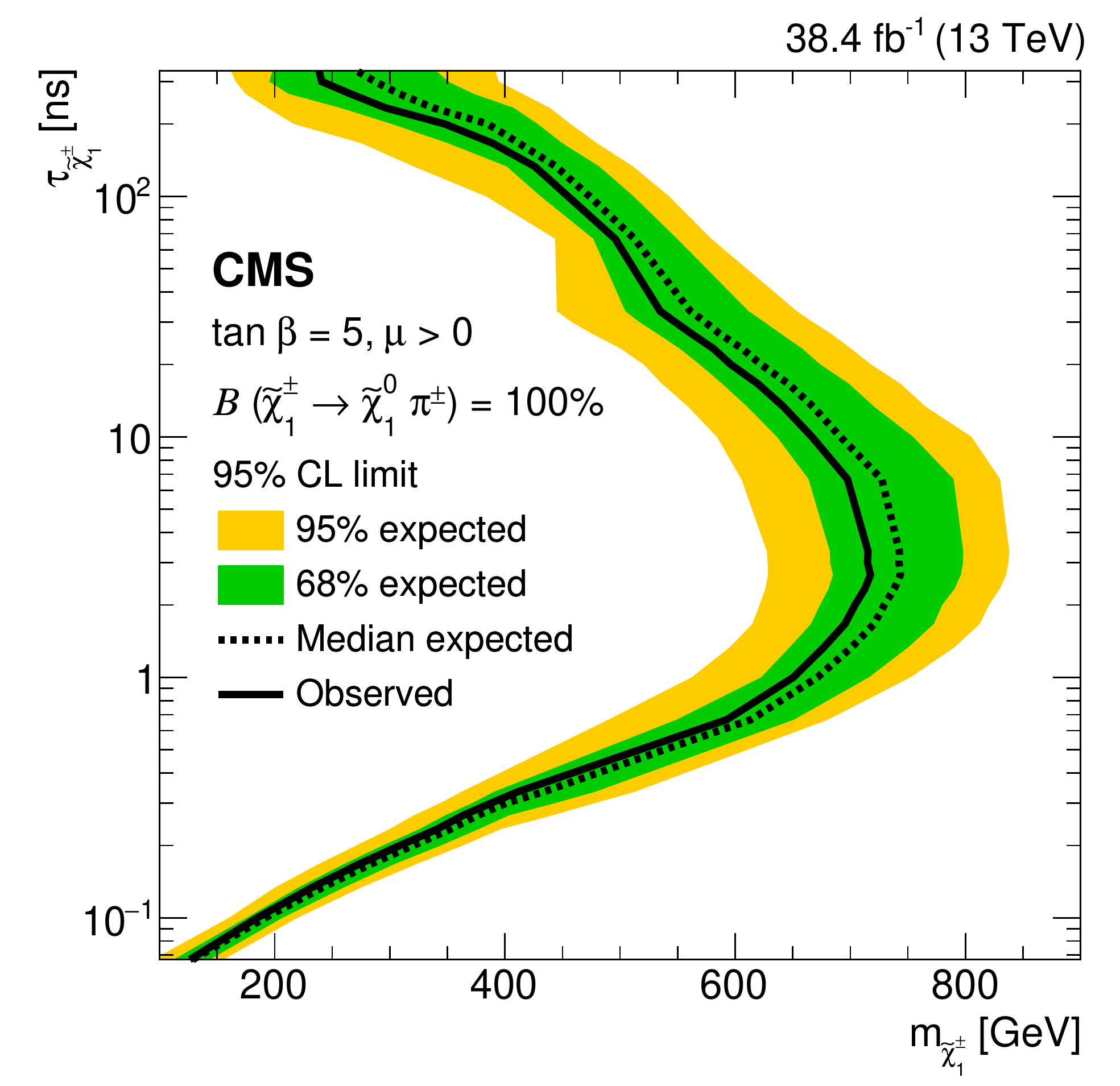}
  \caption{The expected and observed constraints on chargino lifetime and mass.
  The region to the left of the curve is excluded at 95\% \CL.}
  \label{fig:limitsLifetimeVsMass}
\end{figure}

\begin{figure}[htbp]
  \centering
  \includegraphics[width=0.6844\textwidth]{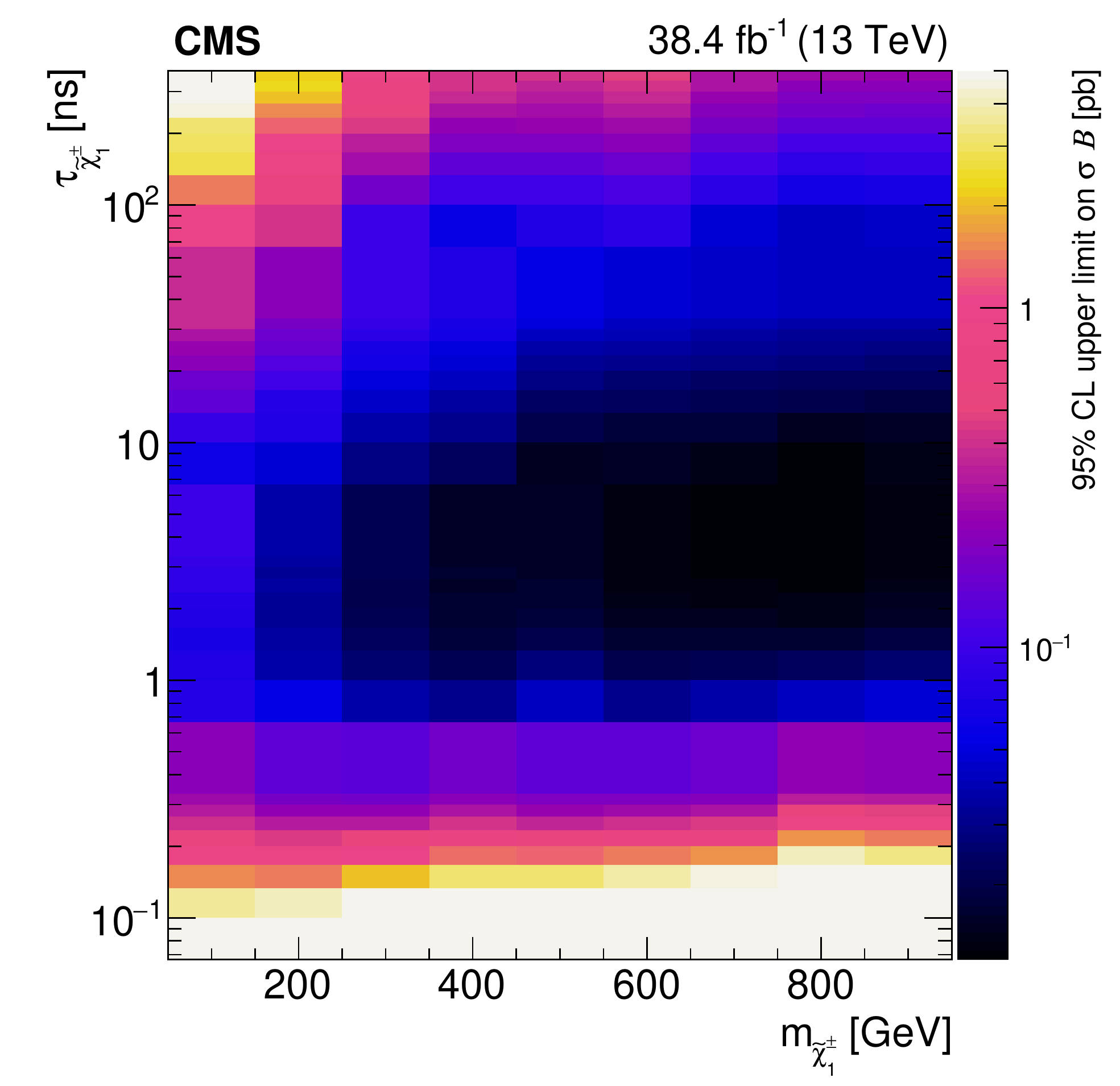}
  \caption{The observed 95\% \CL upper limits on the product of the cross
  section for direct production of charginos and their branching fraction to
  $\PSGczDo\Pgppm$ as a function of chargino mass and lifetime. The direct
  chargino production cross section includes both $\PSGczDo\PSGcpmDo$ and
  $\PSGcpmDo\PSGcmpDo$ production in roughly a 2:1 ratio for all chargino
  masses considered.}
  \label{fig:limitsCrossSectionVsLifetimeVsMass}
\end{figure}

The expected and observed limits on the product of $\sigma$ and $\mathcal{B}$
are shown in Fig.~\ref{fig:limitsCrossSectionVsMass} as a function of chargino
mass, for three different chargino lifetimes. Both $\PSGczDo\PSGcpmDo$ and
$\PSGcpmDo\PSGcmpDo$ production are included in $\sigma$ as a function of
chargino mass as given in theory, which predicts a ratio of roughly 2:1 over
the masses considered. The intersection of the theoretical prediction and the
upper limit on the cross section is used to set a constraint on the mass of the
chargino, for a given chargino lifetime. This procedure is repeated for a large
number of chargino lifetimes, in order to produce a two-dimensional constraint
on the chargino mass and mean proper lifetime, which is shown in
Fig.~\ref{fig:limitsLifetimeVsMass}. Charginos with a lifetime of 3
(7)\unit{ns} are excluded up to a mass of 715 (695)\GeV. Conversely, charginos
with a mass of 505\GeV are excluded for lifetimes from 0.5 to 60\unit{ns}.
Figure~\ref{fig:limitsCrossSectionVsLifetimeVsMass} shows the observed limits
on the product of the cross section for direct production of charginos and
their branching fraction to $\PSGczDo\Pgppm$.

\section{Summary}
\label{sec:summary}

A search has been presented for long-lived charged particles that decay within
the CMS detector and produce the signature of a disappearing track. In a sample
of proton-proton data recorded in 2015 and 2016 at a center-of-mass energy of
13\TeV and corresponding to an integrated luminosity of 38.4\fbinv, seven
events are observed, compared with the estimated background from standard model
processes of $6.5 \pm 0.9 \stat \pm 1.0 \syst$ events. The observation is
consistent with the background-only hypothesis. The results are interpreted in
the context of the anomaly-mediated supersymmetry breaking model, which
predicts a small mass difference between the lightest chargino (\PSGcpmDo) and
neutralino (\PSGczDo). The chargino decays via $\PSGcpmDo \to \PSGczDo \Pgppm$,
and because of the limited phase space available for the decay, the chargino
has a lifetime on the order of 1\unit{ns} and the pion generally has too low
momentum to yield a reconstructed track. If the chargino decays inside the
tracker volume, it can thus produce a disappearing track. We place constraints
on the mass of charginos from direct electroweak production, for chargino mean
proper lifetimes between 0.1 and 100\unit{ns}. Charginos with masses up to 715
(695)\GeV for a lifetime of 3 (7)\unit{ns} are excluded at 95\% confidence
level, as are charginos with lifetimes from 0.5 to 60\unit{ns} for a mass of
505\GeV. These constraints extend the limits set by a previous search for
disappearing tracks performed by the CMS Collaboration~\cite{CMS:2014gxa} and
are complementary to the limits set by searches for heavy stable charged
particles, which exclude charginos with much longer
lifetimes~\cite{Chatrchyan:2013oca,Khachatryan:2015lla}. For chargino lifetimes
above $\approx$0.7\unit{ns}, the present search places the most stringent
constraints using a disappearing track signature on direct chargino production.

\begin{acknowledgments}
We congratulate our colleagues in the CERN accelerator departments for the excellent performance of the LHC and thank the technical and administrative staffs at CERN and at other CMS institutes for their contributions to the success of the CMS effort. In addition, we gratefully acknowledge the computing centers and personnel of the Worldwide LHC Computing Grid for delivering so effectively the computing infrastructure essential to our analyses. Finally, we acknowledge the enduring support for the construction and operation of the LHC and the CMS detector provided by the following funding agencies: BMWFW and FWF (Austria); FNRS and FWO (Belgium); CNPq, CAPES, FAPERJ, and FAPESP (Brazil); MES (Bulgaria); CERN; CAS, MoST, and NSFC (China); COLCIENCIAS (Colombia); MSES and CSF (Croatia); RPF (Cyprus); SENESCYT (Ecuador); MoER, ERC IUT, and ERDF (Estonia); Academy of Finland, MEC, and HIP (Finland); CEA and CNRS/IN2P3 (France); BMBF, DFG, and HGF (Germany); GSRT (Greece); NKFIA (Hungary); DAE and DST (India); IPM (Iran); SFI (Ireland); INFN (Italy); MSIP and NRF (Republic of Korea); LAS (Lithuania); MOE and UM (Malaysia); BUAP, CINVESTAV, CONACYT, LNS, SEP, and UASLP-FAI (Mexico); MBIE (New Zealand); PAEC (Pakistan); MSHE and NSC (Poland); FCT (Portugal); JINR (Dubna); MON, RosAtom, RAS and RFBR (Russia); MESTD (Serbia); SEIDI, CPAN, PCTI and FEDER (Spain); Swiss Funding Agencies (Switzerland); MST (Taipei); ThEPCenter, IPST, STAR, and NSTDA (Thailand); TUBITAK and TAEK (Turkey); NASU and SFFR (Ukraine); STFC (United Kingdom); DOE and NSF (USA).

\hyphenation{Rachada-pisek} Individuals have received support from the Marie-Curie program and the European Research Council and Horizon 2020 Grant, contract No. 675440 (European Union); the Leventis Foundation; the A. P. Sloan Foundation; the Alexander von Humboldt Foundation; the Belgian Federal Science Policy Office; the Fonds pour la Formation \`a la Recherche dans l'Industrie et dans l'Agriculture (FRIA-Belgium); the Agentschap voor Innovatie door Wetenschap en Technologie (IWT-Belgium); the F.R.S.-FNRS and FWO (Belgium) under the ``Excellence of Science - EOS" - be.h project n. 30820817; the Ministry of Education, Youth and Sports (MEYS) of the Czech Republic; the Lend\"ulet ("Momentum") Programme and the J\'anos Bolyai Research Scholarship of the Hungarian Academy of Sciences, the New National Excellence Program \'UNKP, the NKFIA research grants 123842, 123959, 124845, 124850 and 125105 (Hungary); the Council of Science and Industrial Research, India; the HOMING PLUS program of the Foundation for Polish Science, cofinanced from European Union, Regional Development Fund, the Mobility Plus program of the Ministry of Science and Higher Education, the National Science Center (Poland), contracts Harmonia 2014/14/M/ST2/00428, Opus 2014/13/B/ST2/02543, 2014/15/B/ST2/03998, and 2015/19/B/ST2/02861, Sonata-bis 2012/07/E/ST2/01406; the National Priorities Research Program by Qatar National Research Fund; the Programa Estatal de Fomento de la Investigaci{\'o}n Cient{\'i}fica y T{\'e}cnica de Excelencia Mar\'{\i}a de Maeztu, grant MDM-2015-0509 and the Programa Severo Ochoa del Principado de Asturias; the Thalis and Aristeia programs cofinanced by EU-ESF and the Greek NSRF; the Rachadapisek Sompot Fund for Postdoctoral Fellowship, Chulalongkorn University and the Chulalongkorn Academic into Its 2nd Century Project Advancement Project (Thailand); the Welch Foundation, contract C-1845; and the Weston Havens Foundation (USA).
\end{acknowledgments}

\bibliography{auto_generated}
\cleardoublepage \appendix\section{The CMS Collaboration \label{app:collab}}\begin{sloppypar}\hyphenpenalty=5000\widowpenalty=500\clubpenalty=5000\vskip\cmsinstskip
\textbf{Yerevan Physics Institute, Yerevan, Armenia}\\*[0pt]
A.M.~Sirunyan, A.~Tumasyan
\vskip\cmsinstskip
\textbf{Institut f\"{u}r Hochenergiephysik, Wien, Austria}\\*[0pt]
W.~Adam, F.~Ambrogi, E.~Asilar, T.~Bergauer, J.~Brandstetter, E.~Brondolin, M.~Dragicevic, J.~Er\"{o}, A.~Escalante Del Valle, M.~Flechl, R.~Fr\"{u}hwirth\cmsAuthorMark{1}, V.M.~Ghete, J.~Hrubec, M.~Jeitler\cmsAuthorMark{1}, N.~Krammer, I.~Kr\"{a}tschmer, D.~Liko, T.~Madlener, I.~Mikulec, N.~Rad, H.~Rohringer, J.~Schieck\cmsAuthorMark{1}, R.~Sch\"{o}fbeck, M.~Spanring, D.~Spitzbart, A.~Taurok, W.~Waltenberger, J.~Wittmann, C.-E.~Wulz\cmsAuthorMark{1}, M.~Zarucki
% ...
\vskip\cmsinstskip
% ...
1:  Also at Vienna University of Technology, Vienna, Austria\\
% ...
\end{sloppypar}
\end{document}